\documentclass{vgtc}                          % final (journal style)
\usepackage{times}                     % we use Times as the main font
\onlineid{0}

\vgtccategory{Research}

\vgtcinsertpkg

\usepackage{xcolor}
\usepackage[normalem]{ulem}

\newif\ifdiff
\ifdiff
  \newcommand{\added}[1]{\textcolor{blue}{#1}}
  \newcommand{\deleted}[1]{\textcolor{red}{\sout{#1}}}
\else
  \newcommand{\added}[1]{#1}
  \newcommand{\deleted}[1]{}
\fi
    \title{Hindsight: Similarity-Based Analytics for Mars Rover Drive Retrieval}

\author {Luke Fiorante\thanks{These authors contributed equally to this work.} \thanks{e-mail: luke\_fiorante@mde.harvard.edu}\\ %
     \scriptsize Harvard University %
\and Leslie Liu\footnotemark[1] \thanks{e-mail: leslieliu@cmu.edu}\\ %
     \scriptsize Carnegie Mellon University %
\and Adam Xu \thanks{e-mail: axu5@inside.artcenter.edu} \\ %
     \scriptsize ArtCenter College of Design %
\and Xianmei Lei, Darwin Chiu, Krys Blackwood\\ %
      \parbox{2.6in}{\scriptsize \centering Jet Propulsion Laboratory, Caltech}
\and Maggie Hendrie\\ %
     \scriptsize ArtCenter College of Design %
\and Scott Davidoff\\ %
     \scriptsize Space Science Institute %
\and Santiago V. Lombeyda, Hillary Mushkin\\ %
      \parbox{2.6in}{\scriptsize \centering Caltech}
}

\abstract{%
While Mars rover operators plan drives across hazardous Martian terrain and diagnose unexpected faults, the necessary information is distributed across separate systems and often reconstructed through manual correlation and memory. To address this challenge, we partnered with Mars rover operators at the NASA Jet Propulsion Laboratory to introduce \textit{Hindsight}, a visual analytics system that unifies previously disparate rover drive data into a single workspace for search, comparison, and investigation. This paper presents a design study of the Hindsight application. The partnership revealed that operators reason about drives as holistic spatiotemporal episodes rather than discrete parameters. By externalizing operator intuition into an explicit visual query process, we argue that Hindsight transforms analysis into a structured, shareable workflow. \deleted{Preliminary evaluations indicate faster analysis and improved ability to correlate terrain, telemetry, and fault events.}\added{Preliminary feedback from operators suggests Hindsight supports their ability to correlate terrain, telemetry, and fault events within a single workspace.}

}

\keywords{Design study, Visual analytics, Similarity search, Dynamic time warping, Rover operations.}

\teaser{
  \centering
  \includegraphics[width=\linewidth]{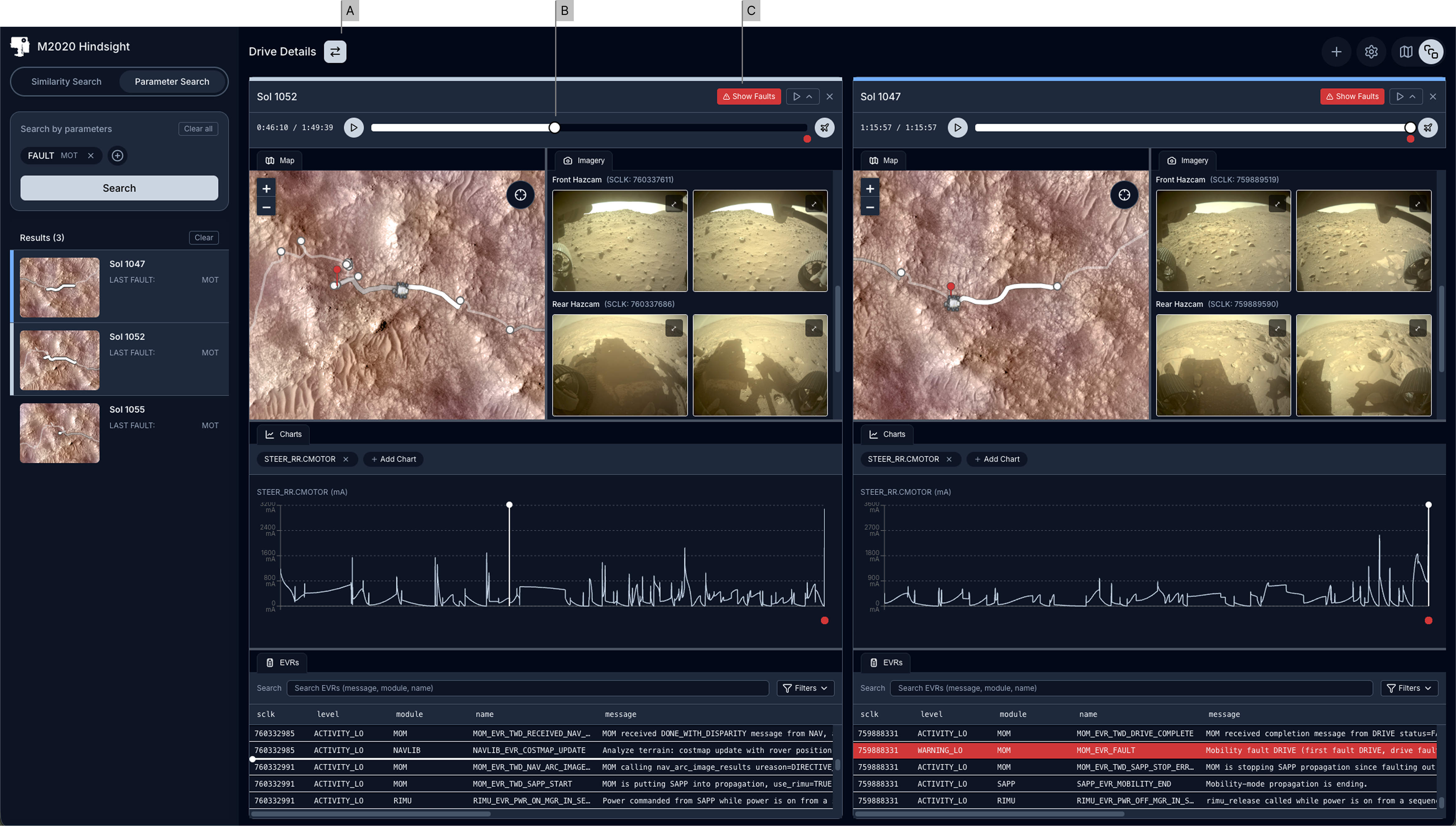}
  \caption{View of the \textit{Hindsight} application: drive panel synchronization (A) ensures consistent arrangement of data sources. The playhead (B) is accessible across the drive panel, tying all visualizations to the same timestamp. The fault in question is highlighted across data in red via the fault overlay (C).}
  \label{fig:fig1}
}
\graphicspath{{figs/}{figures/}{pictures/}{images/}{./}} % where to search for the images

\usepackage{tabu}                      % only used for the table example
\usepackage{booktabs}                  % only used for the table placeholder text
\usepackage{mwe}                       % used to generate placeholder figures
\usepackage{xcolor}
\usepackage{ccicons}                   % package to be able to use icons from creative commons
\usepackage{enumitem}
\usepackage{amssymb} % Required for \geqslant

\usepackage{mathptmx}                  % use matching math font

\begin{document}

%%%%%%%%%%%%%%%%%%%%%%%%%%%%%%%%%%%%%%%%%%%%%%%%%%%%%%%%%%%%%%%%
%%%%%%%%%%%%%%%%%%%%%% START OF THE PAPER %%%%%%%%%%%%%%%%%%%%%%
%%%%%%%%%%%%%%%%%%%%%%%%%%%%%%%%%%%%%%%%%%%%%%%%%%%%%%%%%%%%%%%%

\firstsection{Introduction}

\maketitle
% \renewcommand{\thefootnote}{\fnsymbol{footnote}}
% \footnotetext[1]{These authors contributed equally to this work.}

%  TODO alternative introduction? Santiago's suggestions incorporated
Operating the Perseverance Mars rover is an exercise in extreme risk management~\cite{biesiadecki2005mars,verma2023results}. Rover operators at NASA Jet Propulsion Laboratory (JPL) \deleted{are responsible for }\added{handle }two recurring, high-stakes tasks: 1) planning safe \textit{drives} across Martian terrain that span a Martian day (\textit{sol}) and 2) diagnosing faults when drives end unexpectedly. Both tasks are deeply rooted in historical precedent. The recurring question for operators is, “Have we seen this situation before, and how did we handle it?” Existing operation tools fragment this work across separate single-purpose interfaces, each indexing one slice of a drive's data\deleted{—the telemetry, imagery, paths, or software logs}. To assemble a precedent, operators \deleted{must }look up a past drive by sol number, open each tool in turn, and \deleted{manually }correlate by timestamp. % The implicit assumption is that the data \textit{is} the institutional record and that retrieving the right files gives operators what they need.

To address this analysis gap, we partnered with JPL rover operators over ten weeks to create \textit{Hindsight}, an interactive visual analytics application that unifies historical rover drive data into a single workspace. %and pairs it with a search interface that encodes operators’ analogical language directly. 
This paper presents a design study~\cite{sedlmair2012Design} of the Hindsight application, elaborated through a participatory design~\cite{sanders2008co-creation} engagement with JPL rover operators. Our research argues that knowledge of past drives is stored in two places: in the data, and in operators' episodic memory of how those drives unfolded. Operators do not retrieve drives by parameter thresholds; instead, they retrieve them by resemblance to a remembered episode (“a drive like the one where the rover traversed sand on a slope”). %When asked to describe memorable drives, operators sketched terrain shapes and bird's-eye views of rover paths rather than tabulating values. 
Preliminary evaluation indicates that Hindsight's unified, within-drive view that aligns telemetry, imagery, paths, and software logs around any moment of interest, including faults, better externalizes what operators previously reconstructed from memory across separate tools.

\section{Background} \label{sec:background}

\subsection{Challenges in Rover Operations} \label{sec:rover-operations}

The Perseverance rover’s mission is to seek signs of ancient microbial life and collect rock samples for their eventual return to Earth. To support this objective, JPL rover operators plan and execute rover drive activities to reach science targets across Mars~\cite{verma2022first,verma2023results}. These drive activities are planned at both the strategic level, which focuses on long-term and long-range routing, and the tactical level, which focuses on daily route planning~\cite{verma2022first}. At the strategic level, operators primarily rely on orbital imagery and derived slope maps to plan candidate routes toward regions of scientific interest~\cite{rankin2023perseverance}. This is complemented by onboard imagery from previously driven areas and prior operational experience\deleted{, which together inform the operator’s intuition about traversability in unknown terrain}. At the tactical level, when planning drives through hazardous terrain\deleted{ such as rock, sand, or high-tilt regions}, operators often revisit past drives to identify effective driving strategies\deleted{to understand what worked well and what did not}. 

Due to communication delay between Earth and Mars, there is typically at most one opportunity per sol to dispatch commands. These commands sometimes span multiple sols in a single planning cycle~\cite{leger2005mars}. This limited ability to intervene results in significant operational uncertainty~\cite{verma2022first}. Moreover, driving on Mars remotely is nondeterministic~\cite{biesiadecki2005mars, rankin2020driving, tunstel2005mars}. A drive can proceed as expected, terminate early, or end in a fault that requires diagnosis of the root cause and a recovery~\cite{verma2023results,verma2025robotic}. For this process, rover operators regularly review historical drives to understand past rover behavior, examine faults, and inform future planning~\cite{maimone2026roving}.

Each drive generates rich, multivariate data stored in separate, uncorrelated files, each requiring its own web-based visualization tool, \deleted{which makes it tedious and time-consuming to correlate relevant data}\added{making correlation tedious and time-consuming}~\cite{biesiadecki2008mars}.\deleted{ Given this lack of integrated telemetry visualization, operators spend substantial time aggregating and correlating data into actionable views rather than analyzing them.} Both strategic route planning and fault investigation depend on operators’ intuition, memory, and labor-intensive manual correlation of telemetry and imagery across decoupled tools. This \deleted{creates challenges for }\added{challenges }less-experienced operators, introduces human error, and complicates team communication, as critical knowledge remains largely implicit.\deleted{ As a result, institutional knowledge transfer is limited, leading to inconsistent understanding, duplicated effort, and reduced continuity in decision-making.}

\subsection{Spatial and Temporal Analysis in Operations}

This work participates in a broad discussion at the intersection of robotic operations and visual analytics~\cite{szafir2021connecting} that includes visual-interactive tools for operating individual~\cite{biesiadecki2005mars} and multiple~\cite{rossi2020visual} robots, both via deterministic planning and higher-level autonomy~\cite{alper2019supporting, castano2022operations}. This work draws upon that research, in particular design principles for visual analytics from those high-risk operations contexts~\cite{conlen2018towards}, by foregrounding \deleted{the importance of contextualizing }\added{contextualized }data for time-sensitive decisions and supporting flexible, decision-oriented exploration. Hindsight applies several of these principles to the problem of historical drive retrieval, where operators face a structurally similar challenge\deleted{ involving }\added{: }fragmented tools, manual correlation, and decisions that must be made within a constrained tactical planning window. \added{Hindsight does not introduce new visualization techniques; instead, it couples established representations and interactions and synchronizes access to the full breadth of rover drive data within a single workspace. Its contribution lies in this integration: applying known visual analytics methods, grounded in the direct participation of operators throughout design and implementation, to a fragmented, memory-dependent workflow not previously addressed by an integrated tool.}

These operational challenges are also structurally adjacent to analog research questions in spatial and temporal visualization. Visualization of spatial data has a long history~\cite{Maceachren2001Research}, looking to integrate geospatial information with rich additional dimensions~\cite{Nusrat2016The}, from multimedia ~\cite{Maceachren1994Visualization} to a range of visual encodings ~\cite{Guo2006A}, and their associated computational challenges~\cite{Guo2009Flow}, applying patterns learned to domains from urban mobility~\cite{Beecham2016Map} to geospatial networks~\cite{Schottler2021Visualizing}.

This project in particular draws upon the long methodological history of visualization of time-oriented data~\cite{muller2003visualization}. Wongsuphasawat et al. divide temporal-query interfaces into two broad approaches~\cite{wongsuphasawat2012querying}: \textit{exact match}, in which users specify structured constraints over attributes and time spans, and \textit{similarity search}, in which users provide an example pattern and the system returns records ranked by resemblance. Query-by-example systems have typically been sketch-based and focused on a single variable\deleted{; Wattenberg’s QuerySketch, for instance, matched freehand graphs against stock price histories}~\cite{wattenberg2001sketching}. Correll and Gleicher (2016) later provided a systematic evaluation of matching algorithms for shape-based visual queries, including dynamic time warping, and showed that temporal shape is a more robust basis for perceived similarity than exact value alignment~\cite{berndt1994using,correll2016semantics}. Wongsuphasawat et al. concluded that each paradigm carried complementary strengths and recommended hybrid interfaces that combine them ~\cite{wongsuphasawat2012querying}. Hindsight operationalizes this recommendation\deleted{ in the rover operations domain}, supporting both a parametric mode \deleted{for structured filtering }and a dynamic time warping-based similarity mode\deleted{ that uses either an entire historical drive or a user-defined segment of the rover’s cumulative drive path as the query template. Rather than operating on a single telemetry channel, Hindsight’s similarity search spans a user-selected combination of parameters drawn from the full breadth of rover drive telemetry}.

% telemetry, imagery, paths, flight software logs

% Beyond retrieval, Hindsight’s comparative layout, in which multiple drives are presented side-by-side with synchronized heterogeneous subviews, builds on \textit{coordinated multiple views}, the practice of displaying related data in multiple linked visual representations so that interactions in one view are reflected in all the others, a pattern known as brushing and linking~\cite{roberts2007state}. A layout-synchronization control extends this coordination across panels for cross-drive comparison (Fig.~\ref{fig:fig1}(C)).

\section{Data}
Hindsight integrates several categories of Perseverance mission data. \textit{Temporal data} includes multivariate telemetry time series indexed by \deleted{the rover's }clock time. \textit{Drive telemetry} \deleted{describes both the path driven and characteristics of the traversed terrain, and is used to }\added{includes mechanism positions, motor currents, inertial measurement unit data, and rover pose estimates, used to }monitor mobility mechanism health and diagnose faults. \deleted{This data includes mechanism positions, motor currents, inertial measurement unit data, and rover pose estimates. }\added{Timestamped }\deleted{The temporal dataset also includes timestamped }\textit{flight software logs}\deleted{, which are important to trace progress through sequences of}\added{ trace sequences of} executed commands~\cite{biesiadecki2008mars}. \textit{Spatial data} includes rover paths rendered as waypoints over orbital imagery of Martian terrain. \textit{Visual data} includes timestamped imagery \deleted{taken during the drive as well as the start and end parking positions}\added{captured during the drive and at the start and end parking positions}. We define the \textit{fault context} of a drive as the multimodal slice of these data streams that characterizes a fault at the moment it occurred.

\deleted{The Hindsight prototype was developed with a representative subset of nine mission sols, architected to be able to scale to the full mission archive of over 540 drives across more than 1800 sols (at the time of this writing). This will mean access to over 43 km of traverse, 82 fault diagnoses, over 4 million flight software logs, and 1 TB of elevation maps and terrain imagery.}\added{Hindsight was initially prototyped on a representative subset of nine mission sols and, during deployment at JPL, was expanded to ingest 512 rover drives spanning the mission archive at that time. This larger corpus comprises tens of kilometers of traverse, dozens of fault diagnoses, millions of flight software log entries, and on the order of a terabyte of elevation maps and terrain imagery.}

\section{Design Study} \label{sec:design-study}
Over ten weeks, as part of the Data to Discovery program~\cite{hendrie2022jpl}, we engaged with JPL rover operators, combining contextual inquiry~\cite{beyer1997contextual}, including shadowing operators during active planning sessions and fault investigations in sessions that included participatory design~\cite{sanders2008co-creation}, sketching~\cite{buxton2010sketching}, interactive prototypes, and think-aloud evaluations~\cite{jaspers2004think}. Weekly iterative evaluations enabled iterative refinement of the interface, visual encodings, and search interactions, all implemented in React~\cite{react}. \added{Rather than being specified in advance, Hindsight's specific visual encodings and interactions were co-designed with operators across these iterations, with each revision evaluated against the tasks and reasoning patterns surfaced during field user research.}

A formative design probe \deleted{with rover operators }surfaced the observations that drive Hindsight's design. \deleted{When asked }\added{Asked }to describe memorable past drives, operators \deleted{consistently }produced spatial sketches\deleted{, which included }\added{ --- }annotated terrain diagrams and aerial views of rover paths\deleted{,}\added{ ---} and spoke in analogies (e.g., ``a drive like the one where the rover traversed sand on a slope''), rather than tabulating \deleted{parameter thresholds, which their current tools supported}\added{the parameter thresholds their current tools supported}. While parametric queries capture well-bounded intent cleanly, they require operators to translate \deleted{memories of historical drives}\added{remembered drives} into discrete numerical thresholds. 

\section{User Tasks} \label{sec_user_flow}
Rover operators perform two primary tasks that Hindsight supports, both of which require retrieving, correlating, and comparing historical drive data.

\textbf{Task 1 (T1): Drive Planning.} When planning a drive through challenging terrain (e.g., high slope, sand, rock fields) operators seek past drives under analogous conditions to inform safe planning. As established in Section \ref{sec:design-study}, the query is inherently analogical (e.g., “find drives that looked like this”) and parametric specification is a poor fit. Operators also consult past faults when planning drives over terrain similar to previously problematic areas, making fault investigation and drive planning deeply intertwined.

\textbf{Task 2 (T2): Fault Investigation.} When a drive ends in an unexpected fault, operators must reconstruct the fault context. This aids the search for historically similar drives to assess whether the fault pattern has precedent and how the team responded.

% ====================================================================

\section{The Hindsight Application}
\deleted{Hindsight provides an interactive environment for operators to search, compare, and investigate drives. }To ground the description that follows, we trace a representative drive planning scenario (T1) through the application: an operator is planning Sol 1050, a forward drive through \deleted{a region of }mixed terrain (embedded rock, sand) and needs to identify past drives over comparable conditions\deleted{ to inform a safe and efficient drive}. Sol 1049, executed the previous sol, traversed an extension of the same terrain, making it the operator’s freshest reference point\deleted{. Sol 1049 ended}\added{, and ended} in a fault, meaning the operator also needs to perform a fault investigation (T2). Using Sol 1049 as a query template, the operator searches through the mission archive for additional drives over comparable conditions. \deleted{Section \ref{sec:search-and-similarity} describes how this search unfolds through similarity-based and parametric modes; Section \ref{sec:comparison-and-investigation} describes how coordinated views support detailed investigation of a candidate drive, and side-by-side comparison across drives}

\subsection{Search and Similarity} \label{sec:search-and-similarity}

\begin{figure}[tb]
    \includegraphics[width=\columnwidth]{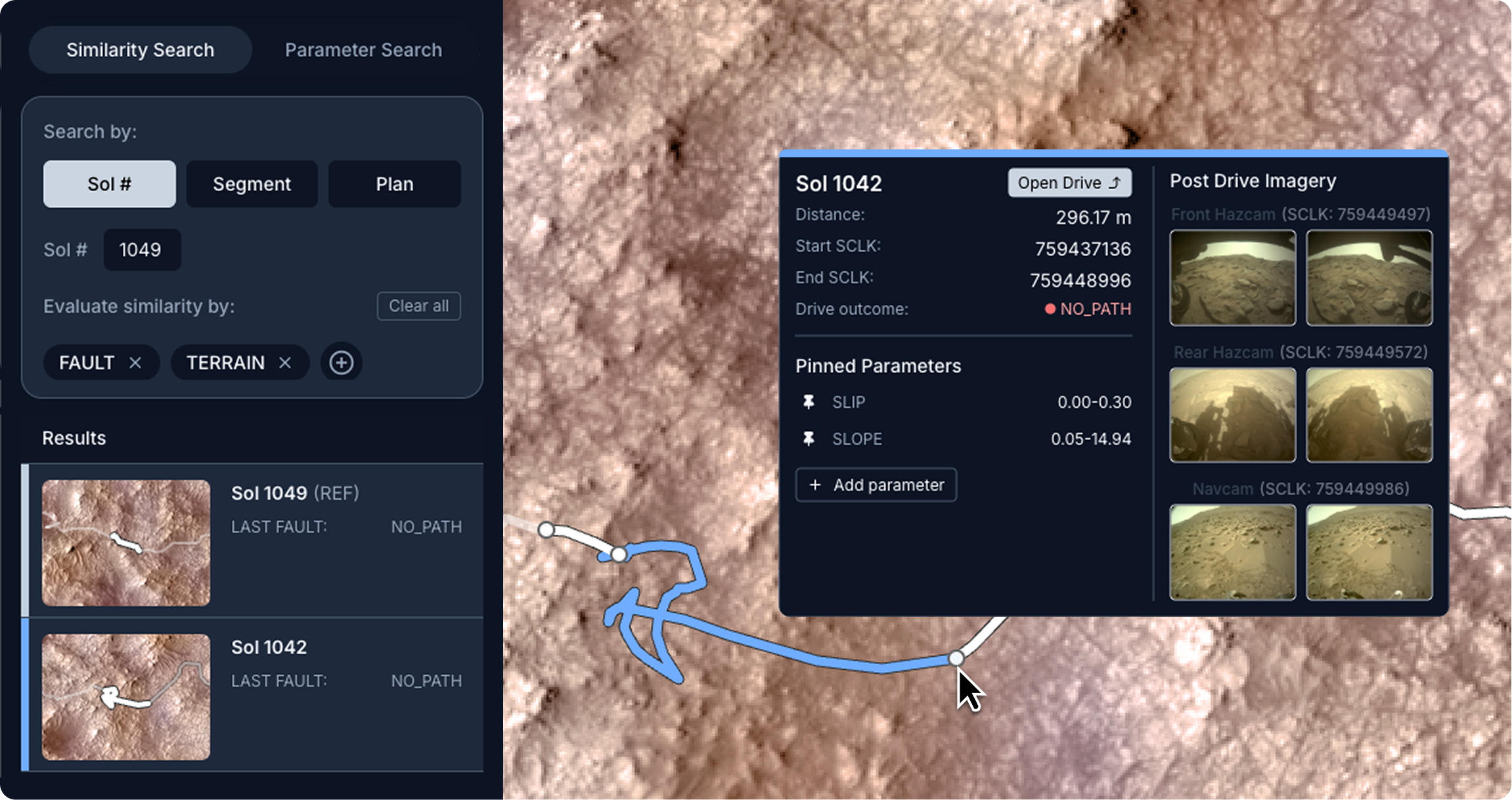} 
        \caption{The operator searches for drives similar to Sol 1049 in fault and terrain type. She inspects the mission archive at a high level by hovering over different drives on the orbital map, then focuses on Sol 1042 in the data summary panel (right) that displays drive statistics.}
    \label{fig:fig2}
\end{figure}

\begin{figure}[tb]
    \includegraphics[width=\columnwidth]{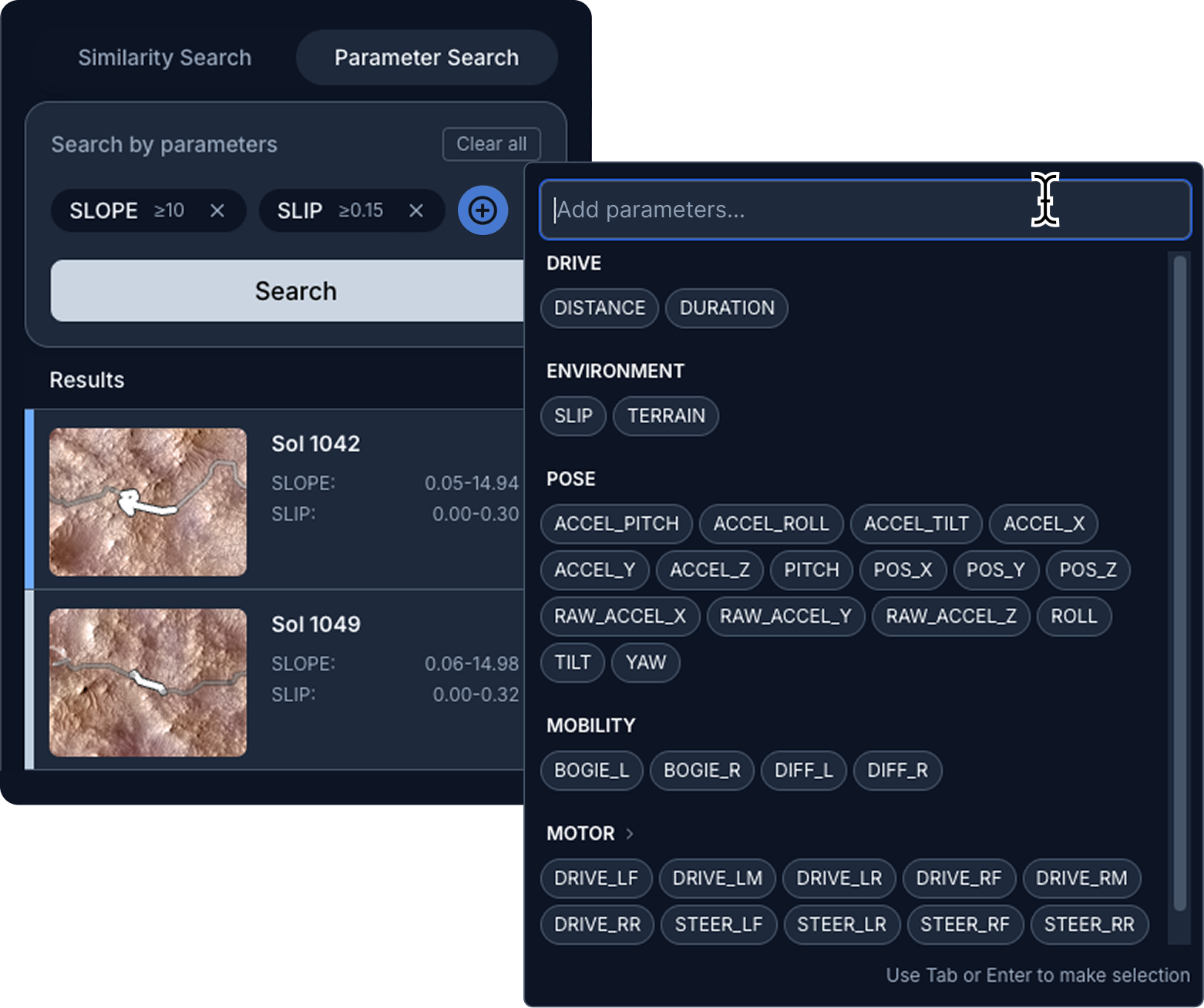}
    \caption{The operator switches to parameter search mode, explicitly defining criteria (here selected $SLOPE\geq10$ and $SLIP\geq0.15$) via the search panel (right), which exposes parameters that correspond to all available data sources.}
    \label{fig:fig3}
\end{figure}

First, the orbital map (Fig.~\ref{fig:fig2}) displays all past mission drives as selectable paths over the Martian surface, letting operators ground a query geographically before specifying its terms. \added{Where drive paths overlap on the map, the topmost is selectable directly; operators can alternatively load any drive by sol number, so map selection is never the sole means of access.} Hovering over any drive displays a preview panel with a compact summary of its key telemetry, imagery, and fault history, allowing operators to build intuition incrementally as they scan. Second, a dual-mode search panel lets operators pose queries either parametrically, by specifying explicit telemetry or terrain conditions, or by similarity, using an existing drive or drive segment as a reference. Search results are returned as ranked list items that, on hover, highlight their corresponding paths on the map\deleted{, closing a visual feedback loop between query and candidate}. Operators may fluidly switch between these search modes\deleted{ in any order}, using results from one to refine parameters in the other\deleted{, progressively externalizing and sharpening their mental model of what constitutes a similar past drive}.

Similarity search is powered by dynamic time warping, aligning sequences that may differ in length or temporal scale\deleted{, a critical capability given that the same maneuver might take different durations across sols due to terrain changes}\added{, since the same maneuver can take different durations across sols as terrain changes}~\cite{berndt1994using}. The algorithm mirrors operators’ thought process: they care about the “shape” of a telemetry signature (e.g., the slip and current pattern of wheels crossing rock at slope), rather than exact timestamp alignment. The \textit{query template} can be either an entire historical drive or a user-defined segment of the rover’s cumulative drive path, which may fall within a single drive or span multiple consecutive drives. \added{To define a segment, an operator clicks along the rover path on the orbital map to place begin and end markers that demarcate the portion of the traverse to use as the query.} This flexibility lets operators isolate specific maneuvers or terrain episodes as the basis for their search, rather than being constrained to whole-drive granularity.

Returning to our scenario, the operator initiates a similarity search using Sol 1049 as the reference and selects the parameters most relevant to planning the next sol’s drive: terrain composition and fault (Fig.~\ref{fig:fig2}). Hindsight returns a ranked list of past drives that most closely match Sol 1049\deleted{ over those parameters}. Hovering each result highlights its corresponding path on the orbital map, letting the operator quickly judge representativeness. Sol 1042 surfaces as a leading candidate.

Parametric search complements this by supporting structured filtering over telemetry thresholds, terrain type, fault presence, and other discrete criteria, with controls exposed across all available data sources (Fig.~\ref{fig:fig3}). \deleted{This search mode handles}\added{It handles} well-bounded questions where operator intent is already expressible as explicit conditions.

\subsection{Comparison and Investigation}  \label{sec:comparison-and-investigation}

\begin{figure}[tb]
    \includegraphics[width=\columnwidth]{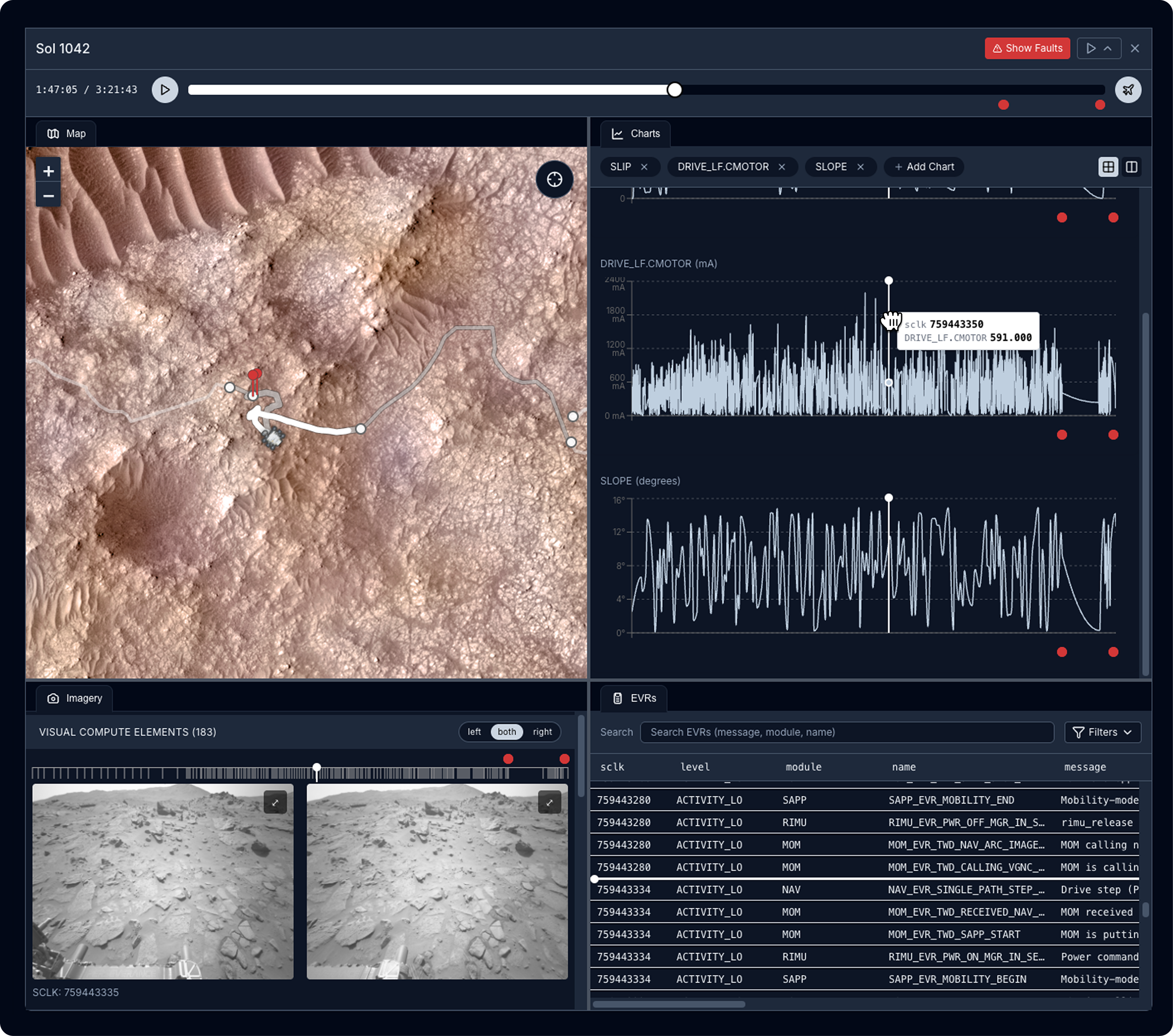}
    \caption{Having narrowed her focus to Sol 1042, the operator explores that Sol’s drive data, scrubbing along rover telemetry charts (right), which synchronizes scrubbing across all other data sources (left, bottom right). Automatic data playback is available via the play button to the left of the synchronized playhead at the top of the drive panel (see also Fig. \ref{fig:fig1}(B)).}
    \label{fig:fig4}
\end{figure}

Investigating drives requires reasoning \deleted{at two levels: }within a single drive and across multiple drives. Hindsight supports both through a unified comparative layout in which each drive is represented by a configurable panel. The drive panel \deleted{is built on }\added{uses }a draggable, splittable window manager with four subviews: an orbital terrain map showing the rover’s path, time-series telemetry charts, flight software logs, and an imagery gallery \deleted{displaying }\added{of }onboard camera captures (Fig.~\ref{fig:fig1}). Up to three drives can be displayed side by side, and a layout synchronization control applies one panel’s configuration to all others\deleted{, ensuring consistent visual comparison across drives}.

Within each drive panel, a synchronized playhead ties all subviews to \deleted{a single }\added{one }point in time, enabling brushing and linking (Fig.~\ref{fig:fig4})~\cite{roberts2007state}. As an operator scrubs through a drive’s timeline, the map updates the rover’s position and traversed path, the telemetry charts highlight the corresponding data values, the imagery panel \deleted{displays the camera frame closest in time}\added{shows the nearest camera frame}, and the flight software logs filter to messages from the same time window. The playhead can be adjusted from the panel-level timeline or from within any subview, with updates propagating across the rest. This synchronization directly addresses the central pain point in the current workflow: the need to \deleted{manually }cross-reference timestamps across \deleted{several }separate tools.\deleted{ An operator can see that a wheel slip event (telemetry) occurred at the same moment the rover encountered a specific obstacle (imagery) and triggered a safety check (flight software logs), all within a single coordinated visualization.}

To support fault investigation (T2), a fault overlay highlights each fault’s occurrence across the relevant subviews: as markers on the drive panel-level timeline, imagery-level timeline, and telemetry charts; as highlighted flight software log lines; and as pinned locations on the orbital map (Fig.~\ref{fig:fig1}(C)). This helps the operator build a complete picture of the fault context.

Returning to our scenario, the operator places Sol 1042 alongside Sol 1049 and aligns their slip and motor current plots: the comparison surfaces where the two drives diverged\deleted{ in performance over comparable terrain} (e.g., the maneuvers in which one drive accumulated more slip than the other). This informs which strategies to replicate or avoid in Sol 1050. The precedent is now directly inspectable rather than dependent on the team’s collective memory of “what worked last time\deleted{,” and the operator can carry that comparison into the planning meeting as concrete visual evidence}.”

\section{Evaluation} \label{sec:evaluation}
\deleted{Hindsight was evaluated iteratively through}\added{We report preliminary results from an iterative, formative evaluation conducted through} weekly think-aloud usability sessions with rover operators, whose input drove refinements to the search interaction, panel layout, and data representation. \added{Three rover operators participated in each weekly session, working through representative tasks they co-defined during our field user research, including locating outlier drives and matching analog errors to prior problems.} We also gathered structured feedback\deleted{ from operators} following hands-on trials.

Operator feedback validated both of Hindsight’s central design commitments. On similarity-based retrieval for fault investigation (T2), one operator noted the tool “would save us so much time when we have a drive fault and want to check whether or not it was similar to a past drive fault,” providing additional context to guide recovery. Regarding the unified workspace, the same operator observed that Hindsight’s synchronized views let operators “explore the telemetry spatially, making it easier to understand how the terrain impacts the telemetry we’re observing.” \deleted{A second operator described their experience as follows:}\added{A second operator offered an unprompted, self-reported impression of time savings, which we report as anecdotal rather than measured:} “I was able to easily pull together different telemetry of interest that I needed to assess a drive in 2–3 minutes, as compared to [the] 15–20 minutes I would have taken [...] using existing operation tools.” 

Operators also suggested potential applicability beyond rover mobility to other rover subsystems and future planetary missions, indicating that Hindsight’s design may generalize to domains where decisions depend on locating analogous episodes in complex operational archives. \added{These findings are preliminary. Our evaluation was formative and embedded within the iterative design process, drawing on a small number of expert operators and self-reported feedback rather than a controlled study. A more comprehensive evaluation, with a larger participant pool and measured task performance across conditions, would be needed to substantiate these observations and is left to future work. Scaling similarity search to the full archive also surfaced performance considerations: a single-variable query over the full corpus ranges from roughly two to several hundred seconds depending on the size of the telemetry variable compared. The upper end of this range is long enough to interrupt an operator's task, so making similarity search consistently performant in live operations remains an objective for future work.}

\section{Conclusion}
Hindsight reframes access to a complex operational archive from a memory-dependent recall task into an explicit visual query workflow. Three design commitments generalize beyond planetary rover operations: a formative design probe can surface the retrieval representation experts actually reason with; similarity-based search can complement parametric filtering when user intent is analogical rather than rule-based; and unifying heterogeneous data streams under a shared temporal reference reduces the cognitive overhead of reconstructing context around a single moment. Future work includes expanding similarity search to use planned drives as queries and capturing operator annotations, building Hindsight into a shared commons of institutional knowledge.

\added{\section*{Supplemental Material}
The Hindsight source code is available at \url{https://github.com/fiorante/hindsight} (archived at \url{https://doi.org/10.5281/zenodo.21115104}). The accompanying data pack is archived at \url{https://doi.org/10.5281/zenodo.21094815}.}

\acknowledgments{
The 2025 Caltech/JPL/ArtCenter Data to Discovery Program and NASA Jet Propulsion Laboratory Mars 2020 funded this project. This research was carried out in part at the Jet Propulsion Laboratory, California Institute of Technology, under a contract with the National Aeronautics and Space Administration (80NM0018D0004). Jenny Rodenhouse from ArtCenter College of Design provided valuable guidance. The Perseverance Rover Operations team and Curiosity (MSL) scientists Ronald S. Sletten and Steve G. Banham offered important insights. \added{In accordance with IEEE policy, we disclose that generative AI tools (Anthropic's Claude Opus 4.1 and Claude Sonnet 4) were used to assist in developing portions of the application's code. The authors reviewed, tested, and take full responsibility for all such code.}
}

\bibliographystyle{abbrv-doi-narrow}

\bibliography{template}
\end{document}